\documentstyle[12pt,epsfig]{article}
\newlength{\dinwidth}                       
\newlength{\dinmargin}                      
 \newcommand{\beq}{\begin{equation}}
 \newcommand{\eeq}{\end{equation}}
 \newcommand{\bea}{\begin{eqnarray}}
 \newcommand{\eea}{\end{eqnarray}}

\begin{document}
\vspace*{1cm}
\begin{center}  \begin{Large} \begin{bf}
    Matching Experimental and Theoretical Jet Definitions
    for Photoproduction at HERA \\
  \end{bf}  \end{Large}
  \vspace*{5mm}
  \begin{large}
    J.M.~Butterworth$^a$, L.~Feld$^b$, M.~Klasen$^c$, G.~Kramer$^c$\\ 
  \end{large}
\end{center}
$^a$ University College London, Physics and Astronomy Dept., London, U.K. \\
$^b$ Physikalisches Institut der Universit\"at Bonn, Bonn, FRG \\
$^c$ II.~Institut f\"ur Theoretische Physik, Universit\"at Hamburg, Hamburg,
FRG \\
\begin{quotation}
\noindent
{\bf Abstract:}
Predictions from a new next-to-leading order (NLO) calculation for direct and
resolved photoproduction of one and two jets are compared to simulated HERA
data. We propose a method to match experimental and theoretical jet definitions
and observe a reduced dependence on jet definitions and hadronization
corrections at larger transverse energies. From the irreducible uncertainty,
we estimate the maximum benefit that can be obtained from increased luminosity
to constrain the structure of the photon and the proton.
\end{quotation}
\section{Introduction}
Hard scattering of real photons off partons can be reliably predicted by
perturbative QCD. The first NLO QCD calculation for complete (direct and
resolved) photoproduction of one and two jets was completed recently
\cite{Kla96}. It is based on the phase space slicing method and uses an
invariant mass cut to integrate soft and collinear singularities analytically.
The program has successfully been tested with an older existing program in
single jet production \cite{Kla93}. In order to be able to compare these
theoretical cross sections to experimental data in photoproduction of jets
from the $ep$ collider HERA \cite{ZEUS,H1}, the soft interaction links between
initial and final state partons and hadrons have to be under control.
Therefore it is only possible to extract new information on the parton
densities in the proton and the photon if the jet definitions in the
measurement and in the theoretical prediction match. In this paper we propose
a method to achieve this for various jet algorithms. The experimental cross
sections are simulated for 1994 HERA conditions, where electrons of energy
$E_e = 27.5$ GeV collided with protons of energy $E_p = 820$ GeV, using
HERWIG 5.8 \cite{Mar92}.
\section{Jet Definitions}
According to the standardization of cone jet algorithms at the Snowmass
meeting in 1990 \cite{Hut92}, calorimeter cells or partons $i$ are included in
a jet, if they have a distance of
\beq
 R_i = \sqrt{(\eta_i-\eta_J)^2+(\phi_i-\phi_J)^2} \leq R \label{eq1}
\eeq
from the jet center and a distance of
\beq
 R_{ij} =\sqrt{(\eta_i-\eta_j)^2+(\phi_i-\phi_j)^2} \leq \frac{E_{T_i}+E_{T_j}}
 {\max (E_{T_i},E_{T_j})} R \label{eq2}
\eeq
from each other. Here, $\eta=-\ln[\tan(\theta/2)]$ is the pseudorapidity
related to the polar angle $\theta$, and $\phi$ is the azimuthal angle.
If two partons have equal transverse energy they may be separated from each
other by as much as $2R$. As parton $j$ does then not lie inside a cone of
radius $R$ around parton $i$ and vice versa, one might with some justification
also count the two partons separately. If one wants to study only the
highest-$E_T$ jet, this ``double counting'' must be excluded. 
The definition of an initiating cluster before a cone is introduced
(``seed-finding'') is not fixed by the Snowmass convention, and different
approaches are possible. The ZEUS collaboration at HERA uses two
different cone algorithms: EUCELL takes the calorimeter cells in a window
in $\eta-\phi$ space as seeds to find a cone with the highest $E_T$. The
cells in this cone are then removed, and the search is continued. On the
other hand, PUCELL was adapted from CDF and starts with single calorimeter
cells. It then iterates cones around each of them, until the set of enclosed
cells is stable. In this case it may happen that two stable jets overlap. 
If the overlapping transverse energy amounts to a large fraction of the jets,
they are merged, otherwise the overlapping energy is split.
In addition, we simulate the same cross sections with the $k_T$ algorithm
KTCLUS \cite{Cat93,Ell93}, where particles are combined if their distance 
\beq
 d_{ij} = \min(E_{T_i},E_{T_j})^2R_{ij}^2
\eeq
is small.
As the same recombination scheme is used, the results are quite similar
to the PUCELL results. In the following we choose $R=1$ throughout.
Partonic jets with a large distance of two
contributing partons are hard to find because of the missing seed in the
jet center. This is especially true for the PUCELL algorithm, which does not
perform a preclustering and does indeed find smaller cross sections and
different hadronization corrections than the less affected EUCELL algorithm.
We propose to model this theoretically by introducing an additional
parameter $R_{\rm sep}$ to restrict the distance of two partons from
each other \cite{Ell92}. This modifies eq.~(\ref{eq2}) to
\beq
 R_{ij} \leq \min\left[\frac{E_{T_i}+E_{T_j}}{\max(E_{T_i},E_{T_j})}R,
 R_{\rm sep}\right].
\eeq
The meaningful range of $R_{\rm sep}$ is between 1 and 2.
For two partons of similar or equal transverse
energies $E_T$, $R_{\rm sep}$ is the limiting parameter, whereas it is the
parton-jet distance $R$ for two partons with large $E_T$ imbalance.
On a NLO three parton final state we find that $R_{\rm sep} = 1.5 ... 2$
for EUCELL and $R_{\rm sep} \simeq 1$ for PUCELL and KTCLUS.

\section{Results}

\begin{figure}[htbp]
 \begin{center}
  \begin{picture}(9,4.5)
   \epsfig{file=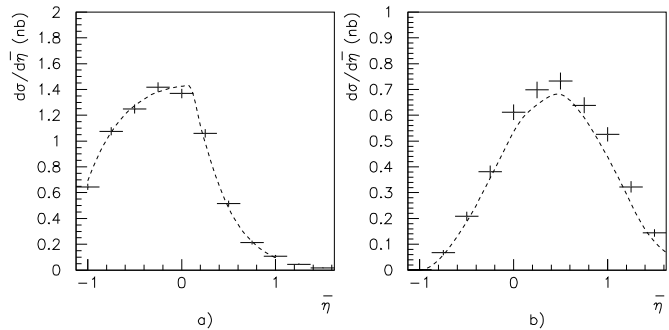,height=4.5cm}
  \end{picture}
  \caption{\label{plot1}{\it Demonstration of compatibility of the HERWIG
  Monte Carlo generator on matrix element level (histograms) and the
  leading-order (LO)
  perturbative QCD prediction (curves) for a) direct and b) resolved dijet
  cross sections} d$\sigma$/d$\overline{\eta}$.}
 \end{center}
\end{figure}
In figure \ref{plot1},
we show the dijet cross section d$\sigma$/d$\overline{\eta}$ with $E_T > 6$
GeV and $\Delta\eta=(\eta_1-\eta_2)\in [-0.5,0.5]$ for a) direct and b)
resolved photoproduction as a function of the average pseudorapidity
of the two observed jets $\overline{\eta} = (\eta_1+\eta_2)/2$. The photon
distribution in the electron is taken from the Weizs\"acker-Williams
approximation with maximum virtuality of $Q_{\max}^2 = 4~\mbox{GeV}^2$ and
longitudinal momentum fraction $y \in [0.2,0.8]$. The direct and resolved
regions are defined by $x_{\gamma}^{\rm OBS} > 0.75$ and
$x_{\gamma}^{\rm OBS} \in [0.3,0.75]$, where the sum in
$x_{\gamma}^{\rm OBS} = \frac{\sum_iE_{T_i}e^{-\eta_i}}{2yE_e}$ runs over
the two jets with largest $E_T$. In both regions, direct and resolved
contributions are added because only their sum is physically meaningful. We
demonstrate the compatibility of the two tools used in this study comparing
the HERWIG Monte Carlo generator on LO matrix element level and the LO
perturbative QCD prediction. We use the CTEQ3L proton and the
GRV$_{\gamma}$(LO) photon structure functions. It was not possible
to calculate $\alpha_s$ in 1-loop approximation in HERWIG, so we took the
(inconsistent) choice of the 2-loop formula for $\alpha_s$ with
$\Lambda_{\mbox{QCD}}^{(4)} = 177$ MeV for this part
in the calculation as well. The scales in HERWIG could also not be changed
from $\mu^2=M_{\gamma}^2=M_p^2=2stu/(s^2+t^2+u^2)$, but the effect with
respect to using $E_T^2$ as in the calculation is very small. After these
adjustments, HERWIG agrees with the LO QCD prediction. \\

\begin{figure}[htbp]
 \begin{center}
  \begin{picture}(10,8)
   \epsfig{file=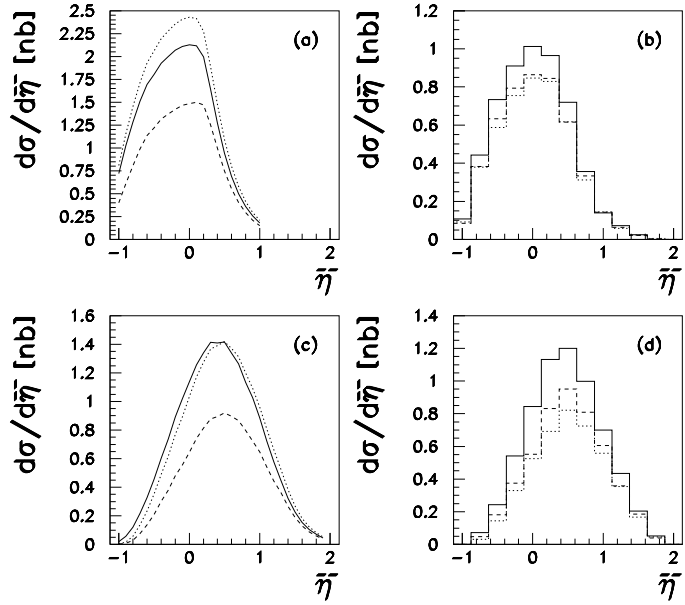,height=8.5cm}
  \end{picture}
  \caption{\label{plot2}{\it NLO (left) and HERWIG (right) predictions for
  direct (top) and resolved (bottom) dijet cross sections}
  d$\sigma$/d$\overline{\eta}$. {\it
  We compare jet double counting (dotted), no $R_{\rm sep}$ (full), and
  $R_{\rm sep}=1$ (dashed) curves with EUCELL (full), PUCELL (dashed),
  and $k_T$ (dotted) histograms.}}
 \end{center}
\end{figure}
Figure \ref{plot2}
shows the same cross section for different jet algorithms.
We compare our new NLO calculation with jet double counting, without
jet double counting, and with $R_{\rm sep} = 1$ to simulated data
from HERWIG with the EUCELL, PUCELL, and KTCLUS algorithms run on the
final state particles. In the
calculation, we now use a NLO set of input parameters,
i.e.~CTEQ3M proton, GRV$_{\gamma}$(HO) photon structure functions, and
2-loop $\alpha_s$ with $\Lambda_{\mbox{QCD}}^{(4)} = 239$ MeV. For the Monte
Carlo, we now take HERWIG including parton showers and hadronization. Due to
the different parameters used in the NLO calculation and in HERWIG, we do not
expect the overall normalization of HERWIG and NLO QCD to agree. However, the
relative changes between no $R_{\rm sep}$ (equivalent to $R_{\rm sep}=2$)
and $R_{\rm sep} = 1$ on the
theoretical side and EUCELL and PUCELL or KTCLUS on the experimental side
show the expected similar behaviour,
so that the $R_{\rm sep}$ parameter is well suited to
match theoretical and experimental jet definitions. Jet double counting does
not correspond to an experimental situation and is only shown to illustrate
its effect on the theory. \\

\begin{figure}[htbp]
 \begin{center}
  \begin{picture}(10,4.5)
   \epsfig{file=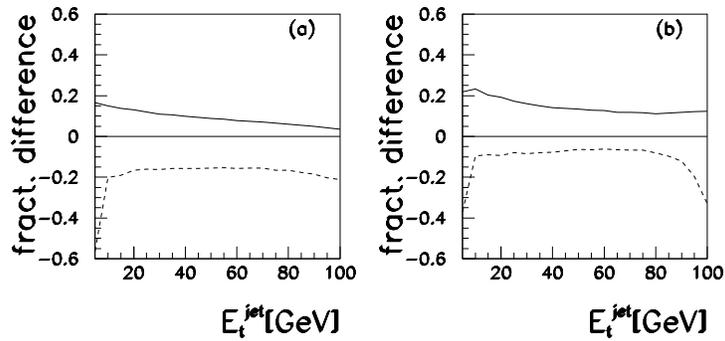,height=4.5cm}
  \end{picture}
  \caption{\label{plot3}{\it $E_T$-dependence of the Snowmass jet definition
  uncertainties for a) direct and b) resolved dijet cross sections}
  d$\sigma$/d$E_T$. {\it We demonstrate the effects of jet double counting
  (full curves) and setting $R_{\rm sep}=1$ (dashed curves) compared
  to no double counting with $R_{\rm sep}=2$.}}
 \end{center}
\end{figure}
The $E_T$-dependence of the Snowmass jet definition uncertainties is shown
in figure \ref{plot3}
for the a) direct and b) resolved dijet cross sections d$\sigma$/d$E_T$, where
we integrated over the complete $\overline{\eta}$ range and over $\Delta\eta\in
[-0.5,0.5]$. The fractional difference of jet double counting from no double
counting amounts to $\sim 20\% $ at 5 GeV and decreases continuously towards
larger $E_T$. Including the parameter $R_{\rm sep} = 1$ lowers the
cross sections by as much as $\sim 40\% $ at 5 GeV, but its influence drops
rapidly and gives a constant difference of about $20 \% $ (direct) and
$10 \% $ (resolved) almost over the whole $E_T$-range. Only at the boundary of
phase space at very large $E_T$, the fractional difference increases again. 
Thus, one should match the jet definitions at small and large $E_T$ even more
carefully. \\

\begin{figure}[htbp]
 \begin{center}
  \begin{picture}(12,6)
   \epsfig{file=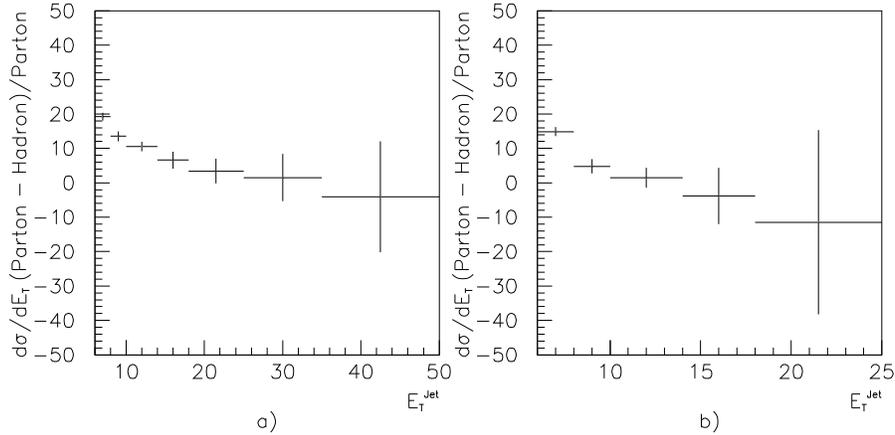,height=6cm}
  \end{picture}
  \caption{\label{plot4}{\it $E_T$-dependence of hadronization corrections for
  a) direct and b) resolved dijet cross sections d$\sigma$/d$E_T$.}}
 \end{center}
\end{figure}
The $E_T$-dependence of the hadronization corrections is shown in figure
\ref{plot4}
for the a) direct and b) resolved dijet cross sections d$\sigma$/d$E_T$. We
integrated again over the complete $\overline{\eta}$ range and over
$\Delta\eta\in[-0.5,0.5]$. At low transverse energies of $E_T\simeq 5$ GeV,
the hadronization corrections amount to $\sim 20\%$ -- a comparable effect
to the theoretical uncertainties discussed before. They decrease very nicely
towards larger $E_T$ and vanish at 30 to 40 GeV to establish the correspondence
between partonic and hadronic jets there. The error bars are due to limited
statistics and could be drastically reduced with more computer time. \\

\begin{figure}[htbp]
 \begin{center}
  \begin{picture}(12,6)
   \epsfig{file=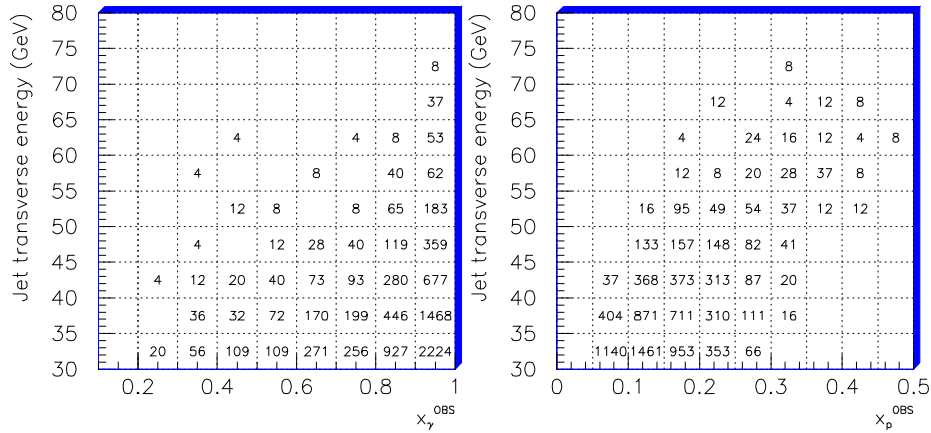,height=6cm}
  \end{picture}
  \caption{\label{plot5}{\it Number of events produced at an increased HERA
  luminosity of 250 pb$^{-1}$ as a function of $x_{\gamma}^{\rm OBS}$
  (left), $x_p^{\rm OBS}$
  (right), and transverse jet energy $E_T$.}}
 \end{center}
\end{figure}
Finally, we estimate the benefit that can be obtained from an increased
HERA luminosity of 250 pb$^{-1}$ to constrain the photon and the proton
parton densities. Figure \ref{plot5}
gives the number of jets produced as a function of $x_{\gamma}^{\rm OBS}$
(left), $x_p^{\rm OBS}$ (right), and transverse jet energy $E_T$, where the
bin sizes reflect the approximate expected experimental resolution. The cuts
applied to the dijet cross section are $y\in [0.2,0.9]$, $E_T > 30$ GeV, and
$\eta < 2$. If we require at least 100 events, where statistical and systematic
errors start to be of comparable size, jets with transverse energies up to
55 GeV can be
measured, where the jet double counting uncertainties and hadronization
corrections are very much reduced. We can still test the photon structure at
large $E_T$ in the region of $x_{\gamma}^{\rm OBS}=0.4 - 1$ and the
proton structure in the region $x_p^{\rm OBS}=0.05 - 0.3$.

\section{Conclusions}
Constraining the proton and photon structure functions in photoproduction of
jets at HERA requires a good understanding of the jets in experiment and in
theory. We used a new NLO calculation and simulated HERA data to match
different experimental jet definitions (EUCELL, PUCELL, and KTCLUS) with
theory predictions with different values for the $R_{\rm sep}$ parameter.
At larger transverse energies, the uncertainties from different theoretical jet
definitions and hadronization corrections are reduced. These regions can be
studied if the HERA luminosity is increased to 250 pb$^{-1}$, thus providing
valuable information on the proton and photon structure functions over large
$x$ ranges.


\begin{thebibliography}{99}
\bibitem{Kla96}
M.~Klasen, G.~Kramer, to be published.
\bibitem{Kla93}
M.~Klasen, G.~Kramer, S.G.~Salesch, Z.~Phys.~{\bf C68} (1995) 113.
\bibitem{ZEUS}
M.~Derrick et al., ZEUS collaboration, Phys.~Lett.~{\bf B348} (1995) 665.
\bibitem{H1}
S.~Aid et al., H1 collaboration, Z.~Phys.~{\bf C70} (1996) 17.
\bibitem{Mar92}
G.~Marchesini et al., Comp.~Phys.~Comm.~{\bf 67} (1992) 465.
\bibitem{Hut92}
J.E.~Huth et al., Proc.~of the 1990 DPF Summer Study on High Energy
Physics, Snowmass, Colorado, edited by E.L.~Berger, World Scientific,
Singapore (1992) 134.
\bibitem{Cat93}
S.~Catani, Yu.L.~Dokshitzer, M.H.~Seymour and B.R.~Webber, Nucl.~Phys.~{\bf
B406} (1993) 187.
\bibitem{Ell93}
S.D.~Ellis, D.E.~Soper, Phys.~Rev.~{\bf D48} (1993) 3160.
\bibitem{Ell92}
S.D.~Ellis, Z.~Kunszt, D.E.~Soper, Phys.~Rev.~Lett.~{\bf 69} (1992) 3615.

\end{thebibliography}
\end{document}